\documentclass[aps,preprint,showpacs]{revtex4}
\usepackage{graphicx}
\usepackage{bm}
\usepackage{subfigure}
\usepackage{amsmath}

\newcommand{\pr}{Phys. Rev.\ }

\newcommand{\jpb}{J. Phys. B\ }
\newcommand{\etal}{{\em et al. }}

\newcommand{\UQ}{School of Mathematics and Physics, University of Queensland, Brisbane, 
QLD 4072, Australia.}

\begin{document}

\title{Asymmetric Gaussian harmonic steering in second harmonic generation}

\author{M.~K. Olsen}
\affiliation{\UQ}
\date{\today}

\begin{abstract}

Intracavity second harmonic generation is one of the simplest of the quantum optical processes and is well within the expertise of most optical laboratories. It is well understood and characterised, both theoretically and experimentally.
We show that it can be a source of continuous variable asymmetric Gaussian harmonic steering with fields which have a coherent excitation, hence combining the important effects of harmonic entanglement and asymmetric steering in one easily controllable device, adjustable by the simple means of tuning the cavity loss rates at the fundamental and harmonic frequencies. We find that whether quantum steering is available via the standard measurements of the Einstein-Podolsky-Rosen correlations can depend on which quadrature measurements are inferred from output spectral measurements of the fundamental and the harmonic. Altering the ratios of the cavity loss rates can be used to tune the regions where symmetric steering is available, with the results becoming asymmetric over all frequencies as the cavity damping at the fundamental frequency becomes significantly greater than at the harmonic.

This asymmetry and its functional dependence on frequency is a potential new tool for experimental quantum information science, with possible utility for quantum key distribution. Although we show the effect here for Gaussian measurements of the quadratures, and cannot rule out a return of the steering symmetry for some class of non-Gaussian measurements, we note here that the system obeys Gaussian statistics in the operating regime investigated and Gaussian inference is at least as accurate as any other method for calculating the necessary  correlations. Perhaps most importantly, this system is simpler than any other methods we are aware of which have been used or proposed to create asymmetric steering.

\end{abstract}


\maketitle


Second harmonic generation (SHG) is one of the simplest phenomena of nonlinear optics~\cite{refSHG} and has long been known as a source of quantum states of the electromagnetic field. Quantum entanglement in terms of the Einstein-Podolsky-Rosen (EPR) paradox~\cite{EPR}, also called steering by Schr\"odinger~\cite{Erwin1,Erwin2}, is one of the central features which differentiates quantum mechanics from classical physics. It has previously been shown that SHG can be used to produce entangled fields at the two frequencies~\cite{sumdiff,SHGEPR}, later called ``harmonic entanglement" by Grosse~\etal~\cite{harmPK}. 

EPR expressed the original paradox in terms of position and momentum measurements. The essential step in their argument was to introduce correlated (entangled) states of at least two particles which persisted when the particles become spatially separated.
According to EPR,
depending on which property of one group of particles that was measured, a prediction 
with some certainty of the values of physical quantities of the other group of particles could be made. If these properties were represented by noncommuting
operators (such as position and momentum), the Heisenberg uncertainty principle could seemingly be violated.
The EPR conclusion was therefore that the description of physical
reality given by quantum mechanics is not complete.

In this work we use the continuous-variable (CV) characterisation of EPR first put on a mathematical footing by Reid~\cite{Margaret}, using quadrature amplitudes, which have the same mathematical properties as position and momentum.
Reid proposed an optical demonstration of the paradox using nondegenerate parametric 
amplification, subsequently realised experimentally by Ou \etal~\cite{Ou}. It is important to note here that this result was the first experimental demonstration of continuous-variable quantum steering, although the authors did not use that terminology.
Later work on CV entanglement saw the introduction of what have become known as the Duane-Simon criteria~\cite{Duan,Simon}, which provide easily measurable correlations to detect bipartite entanglement, in terms of quadrature variances. More recent work~\cite{HMWvolante,Jonesvolante,Ericvolante}, has revisited the early contributions of Schr\"odinger, putting them on a firm mathematical footing in the modern day language of quantum information theory, and reintroducing the term ``steering". Their work defined a ranking of non-classicality, with states violating Bell inequalities being the most non-classical, and these being a subset of states demonstrating the EPR paradox. In the CV case, these themselves were shown to be a subset of states which demonstrated entanglement according to the Duan-Simon inequalities. 

Wiseman \etal  also noted that the EPR inequalities as defined by Reid had a built in asymmetry and raised the question of whether states exhibiting asymmetric steering could be manufactured in the laboratory.
In the tripartite case and with a restriction to Gaussian measurements, such states had already been predicted and analysed~\cite{Ashsteer3}, using a three-mode extension of the original Reid criteria~\cite{MDR3}. In the bipartite case, they have subsequently been predicted in the process of intracavity sum frequency generation~\cite{SFG}, where bichromatic asymmetric steering was analysed theoretically. Further work entailed such states being predicted from 
the intracavity nonlinear coupler~\cite{Sarahvolante} and measured experimentally using parametric downconversion~\cite{Natphotonics}.  We note here that sum-frequency generation does not have the same experimental history as SHG in quantum optics, mainly having been used for spectroscopy, and has quite different stability properties.
Recent works have developed entropic functions for the detection of steering, which show some promise for further investigations of possible asymmetry~\cite{Steve,James}. In this article we will show that the relatively simple system of intracavity second harmonic generation is a good candidate for the realisation of asymmetric harmonic steering, and that all that is required to see this are different cavity loss rates at each of the fundamental and harmonic frequencies.


The Hamiltonian for the intracavity process which couples electromagnetic fields at frequencies $\omega$ and $2\omega$ is written in the rotating wave approximation as~\cite{DFW}
\begin{equation}
{\cal H} = {\cal H}_{int}+{\cal H}_{pump}+{\cal H}_{bath},
\label{eq:Ham}
\end{equation}
where the interaction Hamiltonian is
\begin{equation}
{\cal H}_{int} = i\hbar\frac{\kappa}{2}\left(\hat{a}^{2}\hat{b}^{\dag}-\hat{a}^{\dag\;2}\hat{b}\right),
\label{eq:Hint}
\end{equation}
the pumping Hamiltonian is
\begin{equation}
{\cal H}_{pump} = i\hbar\left(\epsilon\hat{a}^{\dag}-\epsilon^{\ast}\hat{a}\right),
\label{eq:Hpump}
\end{equation}
and the bath Hamiltonian is
\begin{equation}
{\cal H}_{bath} = \hbar\left(\hat{\Gamma}_{a}^{\dag}\hat{a}+\hat{\Gamma}_{a}\hat{a}^{\dag}+\hat{\Gamma}_{b}^{\dag}\hat{b}+\hat{\Gamma}_{b}\hat{b}^{\dag}\right).
\label{eq:Hbath}
\end{equation}
In the above, $\kappa$ represents the effective $\chi^{(2)}$ coupling strength between the two modes, with $\hat{a}$ being the bosonic annihilation operator for excitations at frequency $\omega$ and $\hat{b}$ annihilating excitations at frequency $2\omega$. The classical amplitude of the pump is $\epsilon$ and the $\hat{\Gamma}_{a,b}$ annihilate bath quanta.

Following the standard procedures and making the Markov approximation for the baths~\cite{DFW,QNoise}, we map the Hamiltonian onto a master equation, followed by a Fokker-Planck equation in the positive-P representation~\cite{P+}, making the correspondences $(\hat{a},\hat{a}^{\dag},\hat{b},\hat{b}^{\dag})\leftrightarrow(\alpha,\alpha^{+},\beta,\beta^{+})$. We subsequently find the stochastic differential equations~\cite{SM} for the four positive-P variables,
\begin{eqnarray}
\frac{d\alpha}{dt} &=& \epsilon-\gamma_{a}\alpha+\kappa\alpha^{+}\beta+\sqrt{\kappa\beta}\;\eta_{1}(t),\nonumber\\
\frac{d\alpha^{+}}{dt} &=& \epsilon^{\ast}-\gamma_{a}\alpha^{+}+\kappa\alpha\beta^{+}+\sqrt{\kappa\beta^{+}}\;\eta_{2}(t),\nonumber\\
\frac{d\beta}{dt} &=& -\gamma_{b}\beta-\frac{\kappa}{2}\alpha^{2},\nonumber\\
\frac{d\beta^{+}}{dt} &=& -\gamma_{b}\beta^{+}-\frac{\kappa}{2}\alpha^{+\;2}.
\label{eq:SDE}
\end{eqnarray}
In these equations the $\eta_{i}$ are real Gaussian noise terms with the correlations $\overline{\eta_{i}(t)\eta_{j}(t')}=\delta_{ij}\delta(t-t')$. As always
with the positive-P, the pairs of field variables ($\alpha$ and $\alpha^{+}$ for example) are not 
complex conjugate except in the
mean of a large number of integrated trajectories.

In order to solve the above system of equations we may either integrate them numerically or, in the region where this is applicable, use a linearised fluctuation analysis around the classical steady-state solutions. We follow the second option here, since this allows us to treat the system as an Ornstein-Uhlenbeck process~\cite{SM}, allowing for particularly easy calculation of the output spectral correlations.
It is well known that a Hopf bifurcation
exists at a critical pumping strength, $\epsilon_{c}=(1/\kappa)\left(\gamma_{b}+2\gamma_{a}\right)\sqrt{2\gamma_{b}(\gamma_{a}+\gamma_{b})}$~\cite{Haken,McNeil}, above which the system enters the self-pulsing regime. A
linearised fluctuation analysis can be performed below this
critical point. To begin calculating the output spectral quantities required, we write the positive-P variables as the sum
of a classical, mean value steady-state part and a fluctuations operator,
e.g. $\alpha=\alpha_{ss}+\delta\alpha$, where $\alpha_{ss}=\overline{\alpha}$ in the steady state.  
We can now write an equation of motion for the vector of fluctuation operators, $\delta\hat{X}=\left[\delta\alpha,\delta\alpha^{+},\delta\beta,\delta\beta^{+}\right]^{\mathrm T}$,
\begin{equation}
\frac{d}{dt}\delta\hat{X} = -A\delta\hat{X}+B\;d\zeta,
\label{eq:OSeqn}
\end{equation}
where $A$ is the steady-state drift matrix, $B$ is a matrix of the steady-state coefficients for the fluctuations and $d\zeta$ is a vector of Wiener increments.
As long as the eigenvalues of $A$ do not have a negative real part, the
solutions will be stable, and we can find the intracavity spectral correlations via
\begin{equation}
S(\omega) = \left(A+i\omega\openone\right)^{-1}D\left(A^{\mathrm T}-i\omega\openone\right)^{-1},
\label{eq:OSspek}
\end{equation}
where $D=BB^{\mathrm T}$. These are then easily converted into spectral results outside the cavity using the input-output relations of Gardiner and Collett~\cite{inout}. For the results we present here, we use $\gamma_{a}=1$ and $\kappa=0.01$, with varying $\epsilon$ and $\gamma_{b}$.


Wiseman and others~\cite{HMWvolante,Jonesvolante} have shown that the violation of the quadrature inequalities defined by Reid~\cite{Margaret} in her EPR work also demonstrates that the phenomenon of steering is present in a continuous-variable system. As these are the inequalities we use, we will outline them here. 

We begin by defining the two quadratures of each electromagnetic field as
\begin{eqnarray}
\hat{X}_{a} &=& \hat{a}+\hat{a}^{\dag},\nonumber\\
\hat{Y}_{a} &=& -i\left(\hat{a}-\hat{a}^{\dag}\right),
\label{eq:quads}
\end{eqnarray}
with similar definitions for the harmonic field, using $\hat{b}$ and $\hat{b}^{\dag}$. The Heisenberg Uncertainty Principle then requires that
\begin{eqnarray}
V(\hat{X}_{a})V(\hat{Y}_{a}) \geq 1,\nonumber\\
V(\hat{X}_{b})V(\hat{Y}_{b}) \geq 1,
\label{eq:HUP}
\end{eqnarray}
where variances are defined such that $V(A)=\langle A^{2}\rangle-\langle A\rangle^{2}$ and $V(A,B)=\langle AB\rangle-\langle A\rangle\langle B\rangle$.
The procedure given by Reid~\cite{Margaret} then allows us to define inferred variances which basically come from Gaussian best inference of the expectation values of either the $\hat{X}$ or $\hat{Y}$ quadratures at one frequency from measurements on the quadratures at the other frequency. By measuring the values at the fundamental we may infer values at the harmonic by
\begin{eqnarray}
V^{inf}(\hat{X}_{b})  &=& V(\hat{X}_{b})-\frac{\left[V(\hat{X}_{a},\hat{X}_{b})\right]^{2}}{V(\hat{X}_{a})},\nonumber\\
V^{inf}(\hat{Y}_{b})  &=& V(\hat{Y}_{b})-\frac{\left[V(\hat{Y}_{a},\hat{Y}_{b})\right]^{2}}{V(\hat{Y}_{a})},
\label{eq:VinfB}
\end{eqnarray}
while an inference of the fundamental via measurements at the harmonic leads to
\begin{eqnarray}
V^{inf}(\hat{X}_{a})  &=& V(\hat{X}_{a})-\frac{\left[V(\hat{X}_{a},\hat{X}_{b})\right]^{2}}{V(\hat{X}_{b})},\nonumber\\
V^{inf}(\hat{Y}_{a})  &=& V(\hat{Y}_{a})-\frac{\left[V(\hat{Y}_{a},\hat{Y}_{b})\right]^{2}}{V(\hat{Y}_{b})}.
\label{eq:VinfA}
\end{eqnarray}
The EPR paradox and hence steering are demonstrated whenever
\begin{equation}
V^{inf}(\hat{X}_{a})V^{inf}(\hat{Y}_{a}) <1,
\label{eq:ingHUPa}
\end{equation}
or
\begin{equation}
V^{inf}(\hat{X}_{b})V^{inf}(\hat{Y}_{b}) <1.
\label{eq:ingHUPb}
\end{equation}
We stress here that demonstration of these inequalities does not actually violate the Heisenberg Uncertainty Principle, but is rather a demonstration of the nonlocality of Quantum Mechanics, as has been described in detail elsewhere. In the case of symmetric steering, both these inequalities would be violated equally by a system, with the actual subsystem being measured being unimportant. However, Wiseman \etal~\cite{HMWvolante} raised the question as to the possibility of asymmetric steering, where the actual subsystem being measured would be crucial to the results. As stated above, there have been several predictions of this asymmetry for Gaussian measurements~\cite{Ashsteer3,SFG,Sarahvolante}, and an experimental demonstration~\cite{Natphotonics}. We will now show that it is also a feature of the simple process of intracavity second harmonic generation.

\begin{figure}
\begin{center}
\includegraphics[width=0.8\columnwidth]{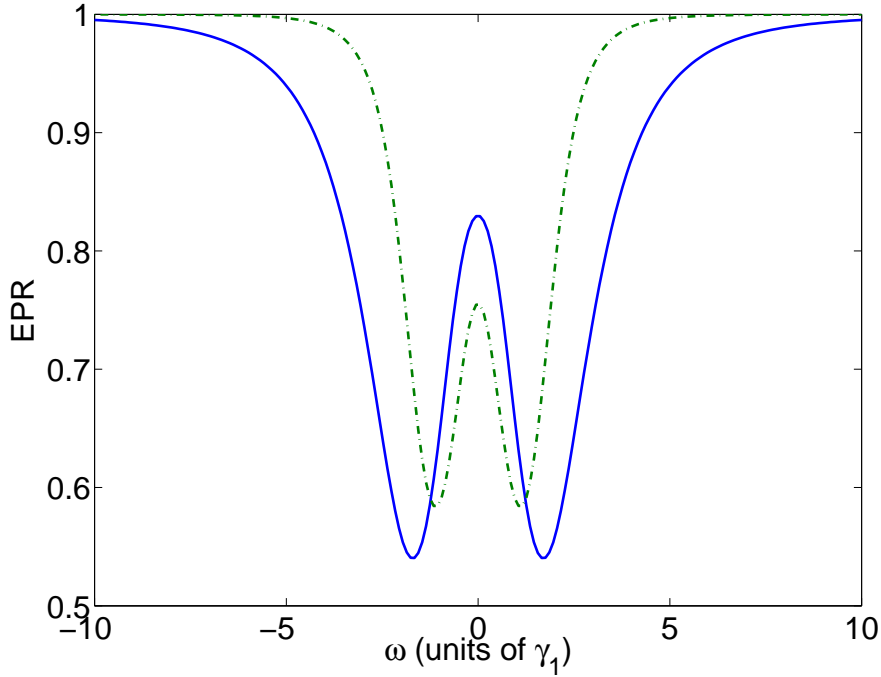}
\end{center}
\caption{(Color online) EPR inferred from fundamental (continuous line) and harmonic (dash-dotted line) for $\gamma_{b}=\gamma_{a}$ and $\epsilon=0.6\epsilon_{c}$. The quantities plotted in this and other graphs are dimensionless.}
\label{fig:equalgamma}
\end{figure}

In intracavity SHG, an obvious and simple source of asymmetry can be introduced by having different loss rates at each frequency for the output mirror. When the mirrors have loss rates such that $\gamma_{b}>\gamma_{a}/2$, we find that steering is possible regardless of which quadratures we use for inference, as shown in Fig.~\ref{fig:equalgamma}. What is apparent, and not seen in totally symmetric systems such as degenerate downconversion, is that the inferred spectral products are not equal, with steering being seen from measurements of the fundamental for frequencies at which it is not seen for measurements of the harmonic. This already allows for a certain degree of asymmetric steering if we restrict ourselves to the appropriate frequency bands.

\begin{figure}
\begin{center}
\includegraphics[width=0.8\columnwidth]{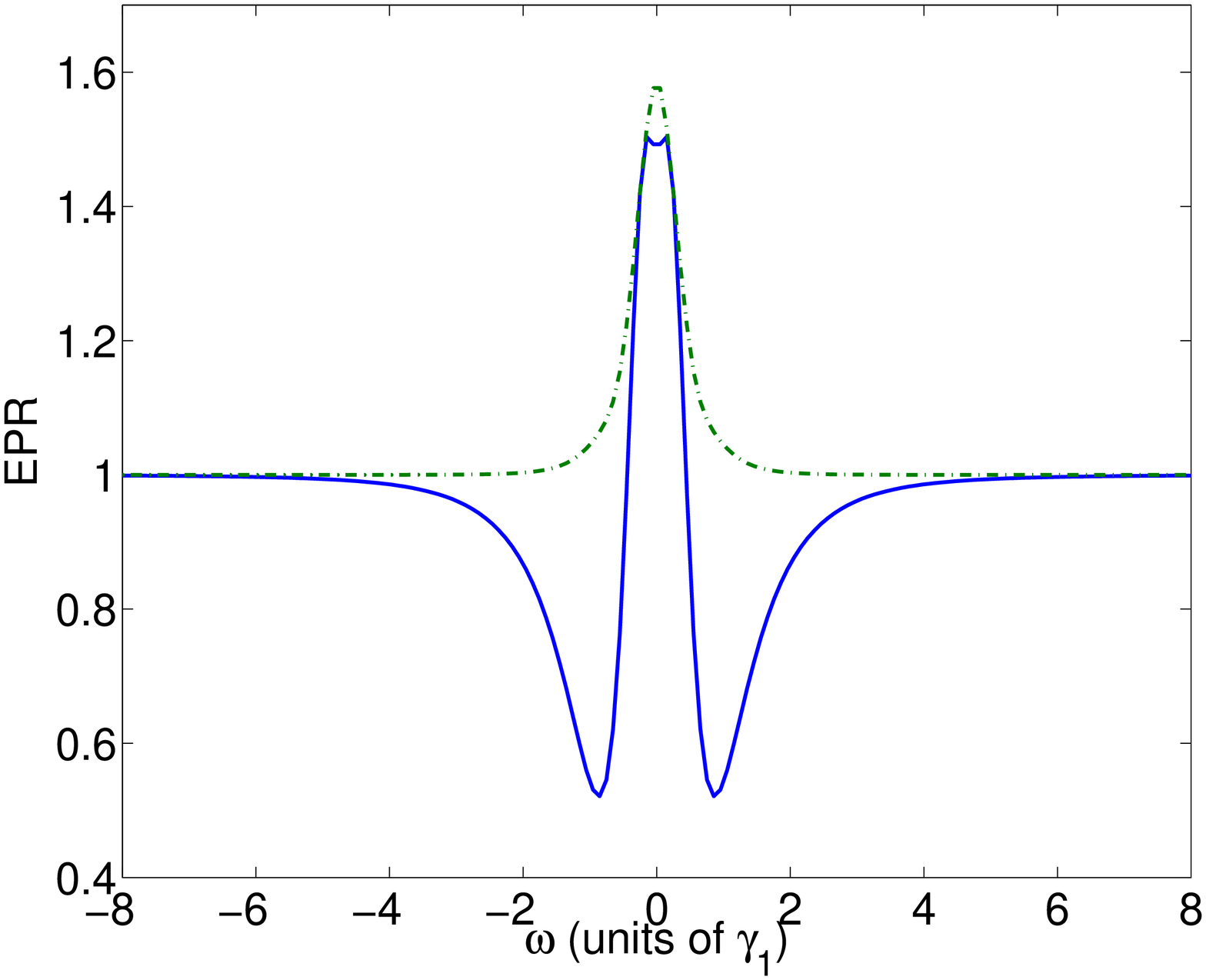}
\end{center}
\caption{(Color online) EPR inferred from fundamental (continuous line) and harmonic (dash-dotted line) for $\gamma_{b}=0.25\gamma_{a}$ and $\epsilon=0.6\epsilon_{c}$.}
\label{fig:harmonic}
\end{figure}

\begin{figure}
\begin{center}
\includegraphics[width=0.8\columnwidth]{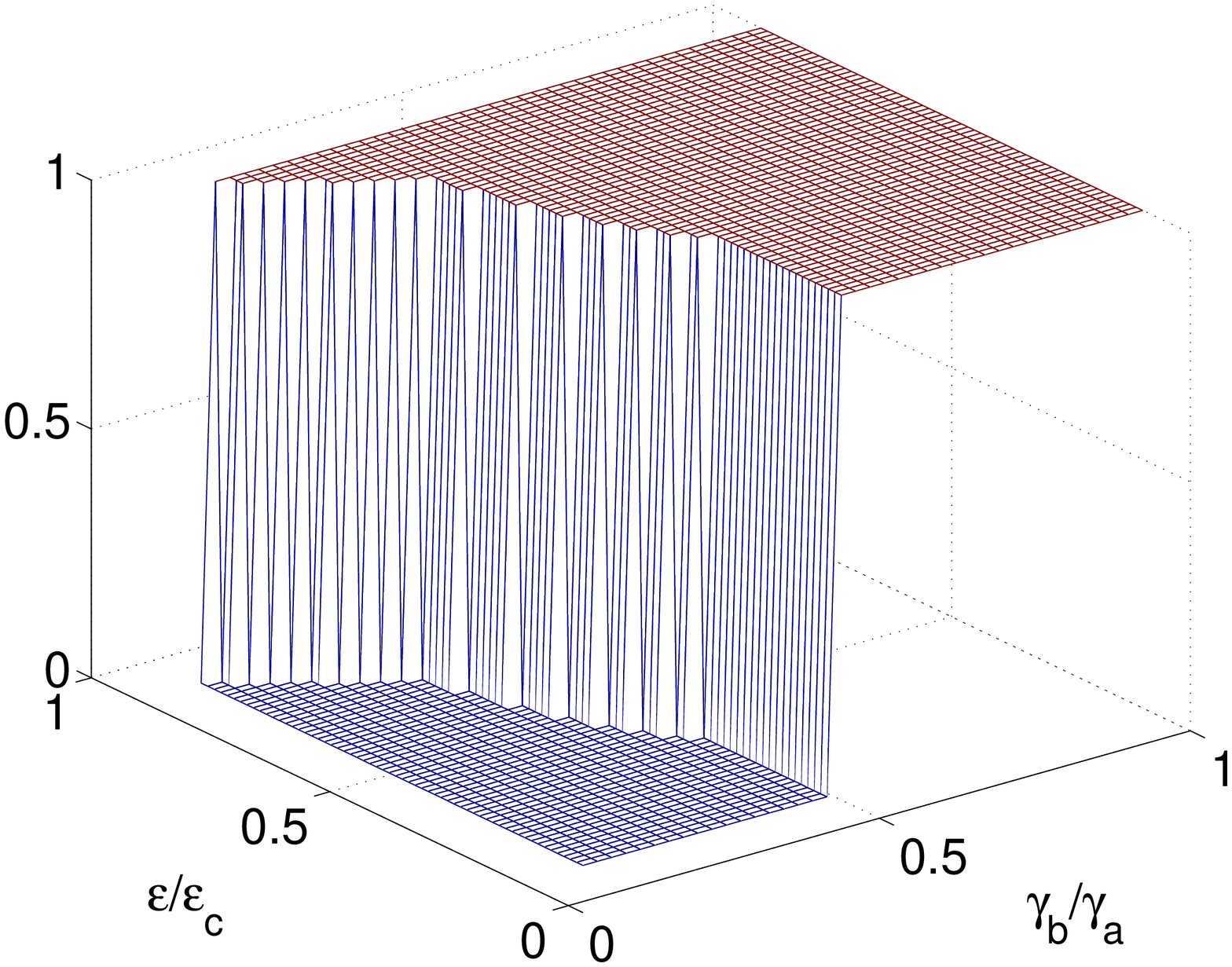}
\end{center}
\caption{(Color online) The presence of asymmetric Gaussian steering at all frequencies is shown here by zero on the vertical axis, as a function of $\gamma_{b}/\gamma_{a}$ and $\epsilon/\epsilon_{c}$. One on the vertical axis shows that there is at least some frequency for which the steering is symmetric.}
\label{fig:canyon}
\end{figure}

However, what we really require is an asymmetry which extends across all frequencies, and we find this as the loss rate at the harmonic decreases. In Fig.~\ref{fig:harmonic} we show the criteria for the same parameters as in Fig.~\ref{fig:equalgamma}, with the only change being to the high frequency cavity loss rate. We readily see that inference using the fundamental allows steering, whereas inference using the harmonic quadratures cannot be used for steering at any frequency. In Fig.~\ref{fig:canyon} we show the presence or absence of asymmetric Gaussian steering across all frequencies as a function of $\gamma_{b}/\gamma_{a}$ and $\epsilon/\epsilon_{c}$, with zero on the vertical axis denoting the presence of total asymmetry and one denoting that there is at least some frequency for which the steering is symmetric. We see that the main criterion is the necessity that the harmonic loss rate be less than the fundamental loss rate, which is experimentally attainable through engineering of the cavity mirrors. This is simpler than the requirements to see asymmetric steering in the nonlinear coupler, previously analysed in ref.~\cite{Sarahvolante}. As stated in our work on the nonlinear coupler, it was necessary to check across all quadrature angles in that system, since the $\chi^{(3)}$ component rotates the Wigner function of the light in quadrature space, but for a resonant $\chi^{(2)}$ system, the violation of the inequalities is greatest for the standard quadrature definitions~\cite{Granja}. We have checked this and it is indeed the case here.

In conclusion, we have shown that asymmetric Gaussian harmonic steering is available in the simple and well characterised system of intracavity second harmonic generation. Importantly, it becomes available by the simple expedient of changing the ratio of the mirror losses at the two frequencies. The entangled outputs are also not squeezed vacuum, as in below threshold downconversion, but have a bright coherent excitation. As well as being of interest from a fundamental point of view, this effect is expected to be of use in quantum communications and quantum cryptography, especially for quantum key distribution~\cite{oneside}. It adds a further tool, and one with which many optical laboratories are familiar, to the techniques available for investigation and use of fundamental quantum mechanics in emerging technologies.

This research was supported by the Australian Research Council under the Future Fellowships Program (Grant ID: FT100100515).

\end{document}